\documentclass{desyproc}

\begin{document}

\title{Characterization of a Transition-Edge Sensor for the ALPS II Experiment}

\author{{\slshape  No\"emie Bastidon$^{1}$, Dieter Horns$^1$, Axel Lindner$^2$,}\\[1ex]
$^1$University of Hamburg, Hamburg, Germany,\\
$^2$Deutsches Elektronen-Synchrotron (DESY), Hamburg, Germany}

\contribID{familyname\_firstname}

\confID{11832}  
\desyproc{DESY-PROC-2015-02}
\acronym{Patras 2015} 
\doi  

\maketitle

\begin{abstract}
	The ALPS II experiment, Any Light Particle Search II at DESY in Hamburg, will look for light (m$< 10^{-4}$ eV) new fundamental bosons (e.g., axion-like particles, hidden photons and other WISPs) in the next years by the
mean of a light-shining-through-the-wall setup. The ALPS II
photosensor is a Transition-Edge Sensor (TES) optimized for $\lambda=1064$~nm photons.
The detector is routinely operated at 80~mK, allowing single infrared photon
detections as well as non-dispersive spectroscopy with very low background
rates. The demonstrated quantum efficiency for such TES is up to~95\% at $\lambda=$1064~ nm as shown in \cite{lita2008}. For 1064 nm photons, the measured
background rate is $< 10^{-2}$ $sec^{-1}$ and the intrinsic dark count rate in a
dark environment was found to be of $1.0\cdot10^{-4}$ $sec^{-1}$
\cite{dreyling2015}.  Latest characterization results are discussed.
\end{abstract}

\section{Single photon detection for ALPS II}
The ALPS II experiment will be looking for new
fundamental bosons. Such a light-shining-through-the-wall experiment requires a
high quantum efficiency low background single-photon detector \cite{Bahre2013}.
A Tungsten Transition-Edge Sensor, which is optimized for low-background 
high quantum efficiency single photon detection, has been developed by NIST
(National Institute of Standards and Technology). 

\section{Detector setup}

\subsection{Tungsten Transition-Edge Sensor}

TESs are superconductive microcalorimeters measuring the temperature difference
$\Delta$T induced by the absorption of a photon. They are operated in a strong
negative electro-thermal feedback corresponding to a constant voltage bias.

When a 1064 nm photon is absorbed by the tungsten chip, the sensor temperature
raises by 0.1 mK. Heating up of the detector brings it from its superconductive stage to close to its normal resistive stage with an increase of the resistance of
$\Delta$R $\approx$ 1 $\Omega$. This leads to a decrease of the current with I
$\approx$ 70 nA. TESs are inductively coupled to a SQUID (Superconducting
Quantum Interference Device) that converts this current variation in a voltage
difference of $\Delta$V $\approx$ -~50~mV.

The ALPS II detector module is constituted of two TESs coupled to a SQUID. Both
detectors are $25\times25~\mu\mathrm{m}^{2}$ large and 20 nm thick. A ceramic standard
mating sleeve towers above each detector, allowing coupling of a standard
single-mode fiber.

\subsection{Adiabatic Demagnetization}

Transition-Edge Sensors are superconductive detectors. The detector needs to be
placed in a bath at $T_{bath}$ = 80 mK $\pm$ 25 $\mu$K. In order to do so, the
TES is placed in an Adiabatic Demagnetization Refrigerator (ADR).

ADR cryostats can reach two low-temperature levels \cite{White2002}. A
temperature baseline of 2.5 K at the colder stages of the cryostat is reached
with the help of a compressor using helium and a pulse-tube cooler. This
cool-down's length in time is only limited by maintenance works and needed
modifications of the setup. Within a cool-down, many phases at 80 mK can be
reached in two hours through adiabatic degmanetization. Such a recharge lasts
approximately 24 hours.

\section{TES Characterization}

\subsection{Pulse shape}

\begin{figure}[hb]
	\centerline{\includegraphics[width=0.5\textwidth]{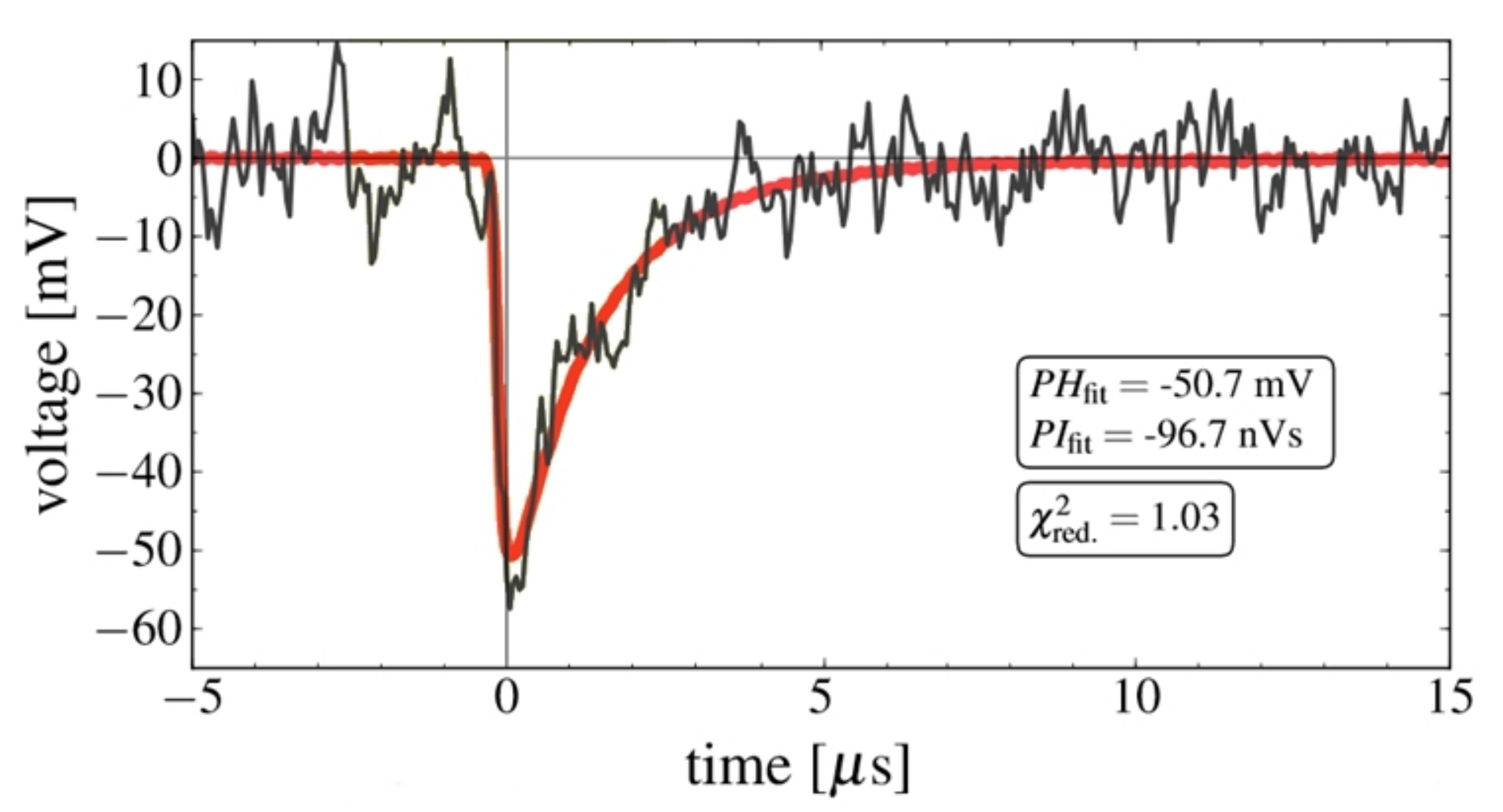}}
	\caption{Infrared single-photon pulse shape.}\label{Fig:pulse}
	\label{sec:figures}
\end{figure}

The average pulse shape for 1064 nm photons shows a Peak Height of PH $\approx$
-50 mV and a Peak Integral of PI $\approx$ -100 nVs (Fig. \ref{Fig:pulse}). A
mask, corresponding to an average pulse, is fitted to the pulses for different
scaling factors \textit{a} and shift values \textit{j} towards the trigger
point \cite{dreyling2015}.

\subsection{Linearity and energy resolution}

\begin{figure}[hb]
	\centerline{\includegraphics[width=0.8\textwidth]{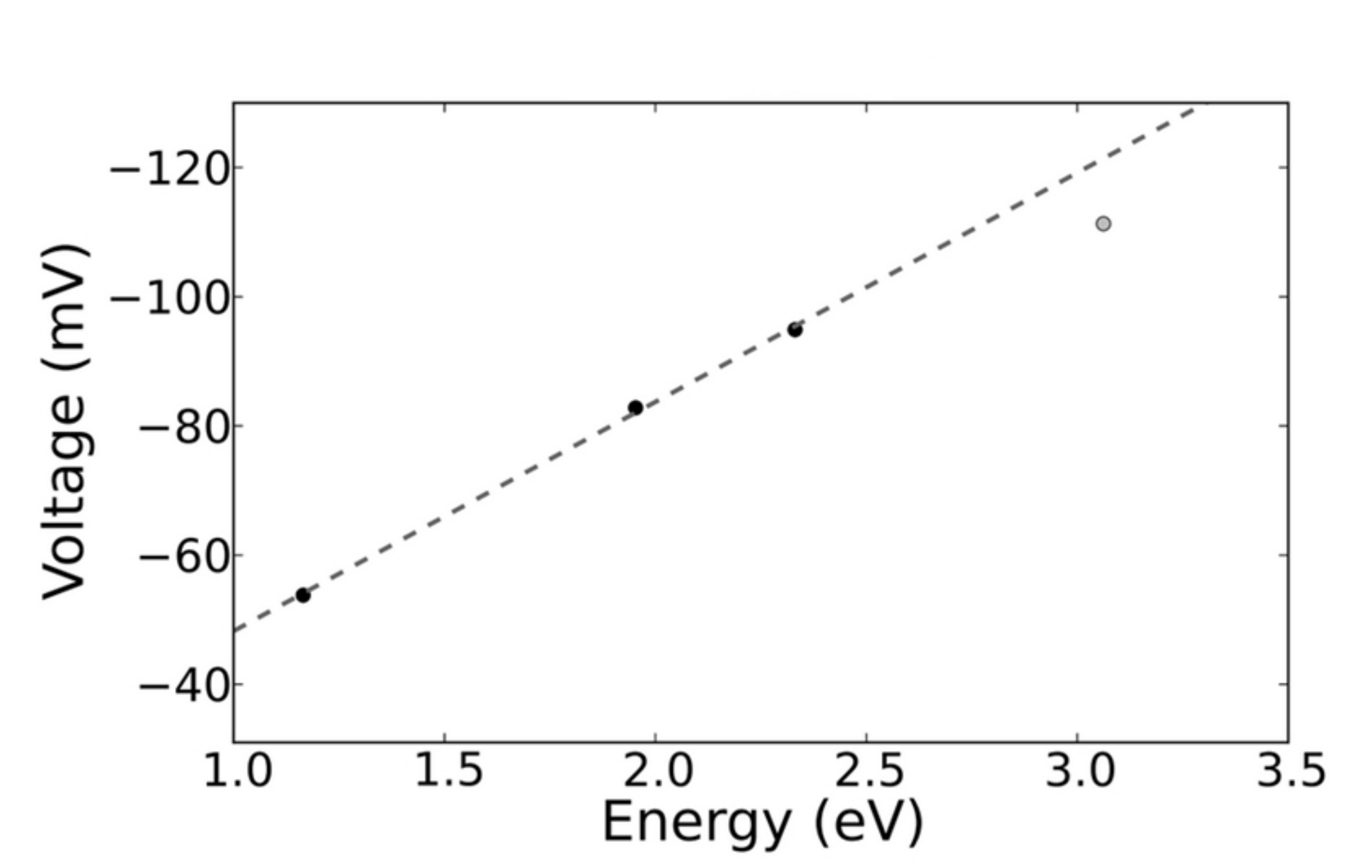}}
	\caption{Average pulse height in units of voltage output as a function of photon energy for the TES. The dashed line is a fit to the first three points.}\label{Fig:linearity}
	\label{sec:figures}
\end{figure}

The linearity of the ALPS II W-TESs was tested by analysing the detector
response to different photon energies. Four different lasers were used to that
purpose (1064, 645, 532, 405 nm). In Figure \ref{Fig:linearity}, the average PH
is shown depending on the energy of the photons absorbed by the detector. The
sensors are linear in our region of interest (1.17 eV) \cite{dreyling2015}. The
non-linearity at higher energies matches expectations (saturation of the
detector).  The energy resolution of the detectors for these different
wavelengths was measured to be $\Delta$E/E $< 8\%$ \cite{dreyling2015}.

\subsection{Stability}

\begin{figure}[hb]
	\centerline{\includegraphics[width=0.8\textwidth]{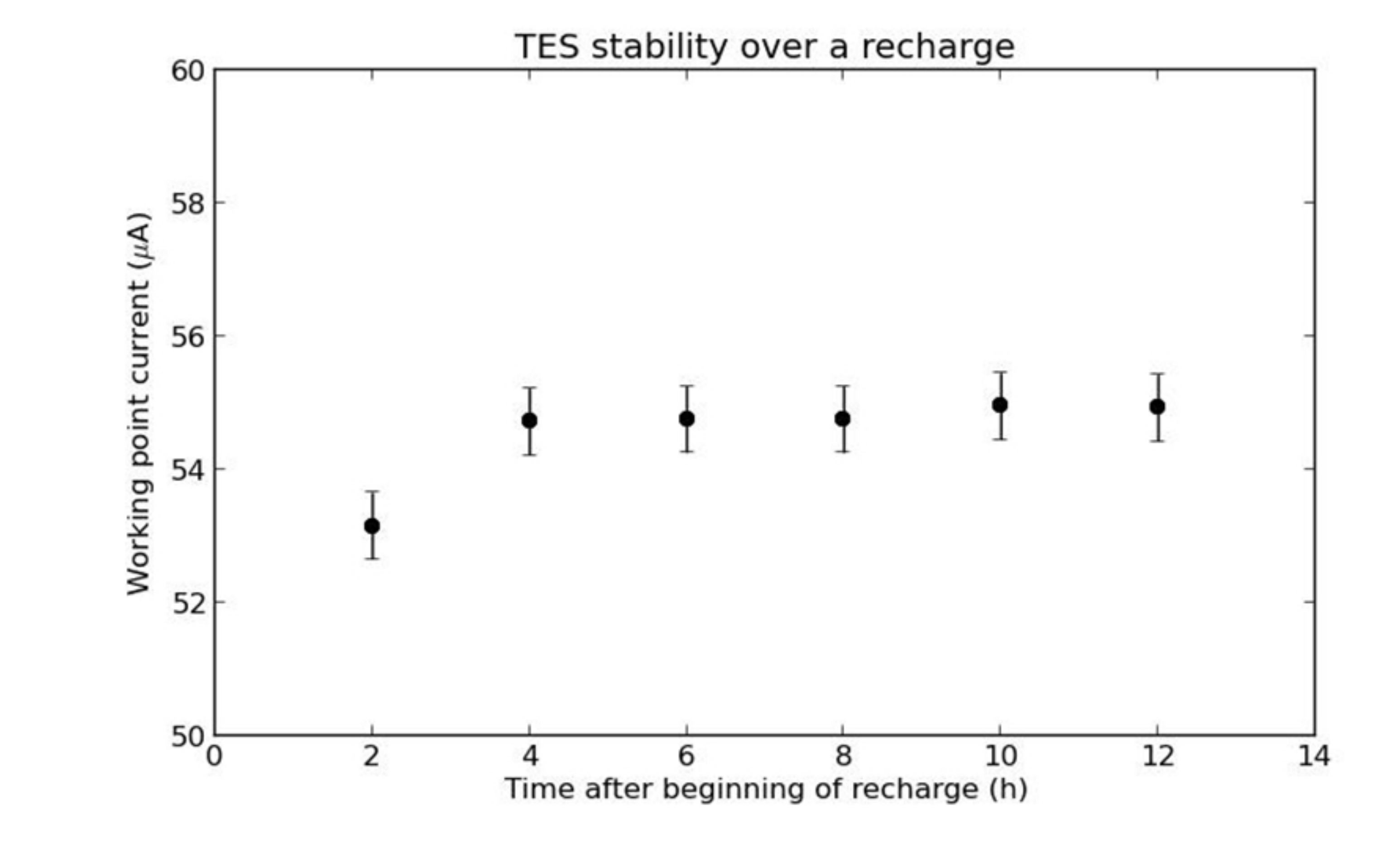}}
	\caption{The TES working point current equivalent to $R_{0}$= 30 \% $R_{normal}$ as a function of time after the beginning of a recharge.}\label{Fig:stability}
	\label{sec:figures}
\end{figure}

Detection stability over time is essential for the ALPS II experiment where
long-term measurements will be performed. Stability during a cool-down as well
as between different cool-downs has been checked successfully. The most
essential characteristic of the detector is its stability during a
recharge-cycle corresponding to the data-taking period. The TES bias current
(i.e. TES working point (Fig. \ref{Fig:stability})) has been measured to be
reasonably stable with a maximum gradient $<$ 1.5 $\mu$A . This variation in
the TES bias current corresponds to a variation in the peak height of
$\Delta$PH $< 3\%$.  Finally, the results have been proven to be operator
independent (adjustment method) \cite{dreyling2015}.

\section{Summary}

Transition-Edge Sensors seem to ideally meet the ALPS II detector challenges.
The characterization of the sensors provided by NIST has demonstrated a good
detector energy resolution as well as a good stability of the pulse shape over
long-term measurements. In addition to this, both detectors have shown a good
linearity in the ALPS II region of interest (1.17 eV).

In the near future, optimization of the detectors quantum efficiency as well as
reduction of the background will be performed.

\section{Acknowledgments}

The authors are grateful to NIST, PTB and Entropy for their technical support.
We would also like to thank J. Dreyling-Eschweiler and F. Januschek.
Finally, we thank the PIER Helmholtz Graduate School for their financial travel support.
 

\begin{footnotesize}

\end{footnotesize}


\end{document}